\documentclass[letterpaper, 10 pt, conference]{ieeeconf}  % Comment this line out if you need a4paper

\IEEEoverridecommandlockouts                              % This command is only needed if 
\usepackage{amsmath} % assumes amsmath package installed
\usepackage{amssymb}  % assumes amsmath package installed
\usepackage{url}
\usepackage{color}

\newcommand{\thm}{\tilde{\theta}}
\newcommand{\ths}{\theta^{\star}}
\newcommand{\T}{\mathsf{T}}
\newcommand{\R}{\mathbb{R}}
\usepackage{bm}                  % \bm{} for bold math

\usepackage{graphicx}
\newtheorem{remark}{Remark}
\newtheorem{problem}{Problem}
\newtheorem{theorem}{Theorem}

\title{\LARGE \bf
Learning reduced-order latent linear models for Kalman filtering of nonlinear systems
}

\author{Manas Mejari, Milad Banitalebi Dehkordi, Dario Piga % <-this % stops a space
\thanks{
This work has been accepted for publication, to appear at the 65th IEEE
Conference on Decision and Control (CDC 2026), Hawaii.
The authors are with IDSIA Dalle Molle Institute for Artificial Intelligence, USI-SUPSI, Lugano, Switzerland.
        {\tt\small email: \{manas.mejari, milad.banitalebi, dario.piga\}@supsi.ch}}%
}

\begin{document}

\maketitle
\thispagestyle{empty}
\pagestyle{empty}

%%%%%%%%%%%%%%%%%%%%%%%%%%%%%%%%%%%%%%%%%%%%%%%%%%%%%%%%%%%%%%%%%%%%%%%%%%%%%%%%
\begin{abstract}
We propose a \emph{filtering-oriented} end-to-end learning framework to identify  reduced-order models explicitly  tailored for state estimation in high-dimensional nonlinear systems. An  \emph{autoencoder} (AE) neural network learns a low-dimensional latent representation of the state together with a lifting map to the original space, while a \emph{reduced-order linear time-invariant} (RO-LTI) model describes the latent dynamics.
The AE and RO-LTI model are trained jointly by minimizing a multi-objective loss that combines reconstruction error with a \emph{filtering objective} based on a differentiable Kalman filter, ensuring that the reduced-order model is  tailored for the downstream state estimation task.
At inference, filtering is performed entirely in the latent space using the  RO-LTI model, and the estimated state is mapped back to the original space via the decoder. Unlike conventional two-stage approaches, in which a reduced-order model is first identified for system approximation and a filter is subsequently designed on top of it, the proposed framework  learns a  \emph{task-oriented} reduced-order model whose parameters are shaped entirely by filtering performance rather than system approximation accuracy alone.
% method explicitly optimizes for filtering performance. 
We further quantify probabilistic bounds on the performance gap between full-order and reduced-order filters using conformal predictions, which do not require assumption on data distribution. The approach is validated on a heat diffusion benchmark, where the full temperature field is reconstructed from sparse measurements.

\end{abstract}

%%%%%%%%%%%%%%%%%%%%%%%%%%%%%%%%%%%%%%%%%%%%%%%%%%%%%%%%%%%%%%%%%%%%%%%%%%%%%%%%
\section{INTRODUCTION}

State estimation for nonlinear dynamical systems is a fundamental problem in control and signal processing. In many applications, including fluid dynamics, power systems, and robotics, the underlying systems are high-dimensional, making real-time state estimation computationally demanding. While the Kalman filter and its variants provide principled estimation frameworks \cite{Sarkka2013,Verhaegen2007book}, their direct application to high-dimensional systems is often impractical in embedded applications due to the computational burden associated with propagating high-dimensional states and covariances.

A common approach to mitigate this challenge is to obtain a reduced-order model (ROM) that captures the dominant system behavior in a low-dimensional space. Classical model reduction techniques, typically classified as SVD-based or Krylov-based methods~\cite{Antoulas2005}, include balanced truncation~\cite{bt1988}, singular perturbation approximation~\cite{LiuAnderson1989},  Hankel norm approximations~\cite{Glover1984} etc.     
However, these approaches are primarily developed for linear systems and typically require explicit system models, limiting their applicability in data-driven settings.

To this end, \emph{data-driven} model reduction methods have gained attention. Such methods include regularization-based system identification approaches~\cite{Bemporad25, forgione24, Pillonetto2016}, the
Loewner framework~\cite{Gosea21}, 
 dynamic mode decomposition~\cite{Proctor2016}, which focus on identifying reduced-order dynamical models from data. Alternatively,  methods such as principal component analysis-based approach~\cite{Annoni2017} and autoencoder-based methods~\cite{KOELEWIJN2024, MASTI2021}  learn   linear or nonlinear embeddings from data which provide compact latent representations of states.  While these approaches have demonstrated promising performance in modeling and prediction, they are typically optimized for data reconstruction or prediction accuracy, and do not explicitly account for state estimation performance. A representation that is optimal for reconstruction or prediction may not be optimal for \emph{filtering}, as it does not account for the structure of the observation model or the noise characteristics of the estimation problem.

For state estimation,  reduced-order filtering techniques have been developed to enable tractable estimation in large-scale systems. For example, in \cite{Farrell2001},  a balanced truncation method is used to obtain a reduced order model which is then employed in a reduced-order Kalman filter.   The contribution~\cite{Nouwens2022} proposes a least-squares kernel Kalman filter that exploits the correlation between state components to obtain a computationally efficient approximate Kalman filter.   However, these methods  follow a two-stage pipeline, where the reduced-order model is learned independently without considering the filtering objective.
 Alternatively, a Monte Carlo–based variant of the Kalman filter, such as the ensemble Kalman filter (EnKF)~\cite{Evensen2003} is developed for high dimensional systems.  In the EnKF,  instead of propagating a full state covariance matrix, a sample covariance is computed from an ensemble of states. But, it  typically requires a large number of samples and is susceptible to sampling errors.

In this work, we propose a \emph{filtering-oriented} learning of reduced-order models of high-dimensional nonlinear systems.  In contrast to Koopman-based approaches~\cite{bevanda2021koopman}, which lift nonlinear dynamics into high-dimensional linear representations for prediction, we take the opposite perspective: we learn a compressed low-dimensional latent space where the dynamics are linear and explicitly tailored for Kalman filtering. This task-oriented representation enables efficient and accurate state estimation in high-dimensional nonlinear systems.

The main idea is to jointly learn a reduced-order LTI model (including its  noise covariances) and  a lifting map to the original state space via an autoencoder.  By embedding the Kalman filter within the learning loop, the model is trained directly with respect  to a filtering objective. The model parameters, covariances and the AE network parameters are learned by minimizing a multi-objective loss  
comprising  reconstruction error,  Kalman filter negative log-likelihood,  decoded state estimation error in the original state space, and a latent alignment term. %In contrast to conventional model-order reduction methods that focus on prediction accuracy, the proposed formulation explicitly optimizes for state estimation performance during training.
This stands in contrast to conventional model-order reduction, where system approximation and state estimation are treated as separate objectives. Here, the two are unified: the reduced-order model is the filtering model, and its parameters are shaped entirely by the estimation task. Overall, this represents a shift from model-centric representations (typically optimized for prediction)  to task-oriented latent representations explicitly designed for filtering.
At  inference, filtering is performed in the reduced-order latent space, and the estimated states are mapped back to the original space via the decoder.  Finally, with conformal predictions,
we provide probabilistic bounds on the performance gap between the full-order Kalman filter and  reduced-order Kalman filters. 

To the best of our knowledge, incorporating the Kalman filter within the training loop has only recently been explored in the \emph{ROAD-EnKF} framework~\cite{Chen2023ROADEnKF}. However, our approach differs in several key aspects. First,~\cite{Chen2023ROADEnKF} assumes a \emph{known} measurement model (including  measurement matrix and noise covariance) of the high-dimensional systems, whereas we learn both the measurement matrix of the latent dynamics and noise covariance directly from data. Second, ROAD-EnKF employs a \emph{decoder-only} architecture, while we use an autoencoder, enabling the encoder to capture \emph{spatial correlations} in the state. Third, our training loss includes a decoded state estimation error in the original state space, in addition to the exact likelihood term, thereby explicitly optimizing for filtering accuracy, while~\cite{Chen2023ROADEnKF} consider only an approximated likelihood term based on ensembles. 
Fourth, ROAD-EnKF relies on EnKF in the latent space, which requires sampling latent states from a prescribed prior for initialization and maintaining an ensemble. This introduces sampling-based approximation errors and makes inference-time complexity dependent on ensemble size. In contrast, our approach leverages a conventional Kalman filter in the latent space and, through the encoder, learns a deterministic mapping from the original state to the reduced-order representation, thereby providing a coherent and data-driven initialization of the latent states without the need for prior sampling.

%Finally, one of our main contributions is providing a detailed theoretical and empirical analysis of probabilistic bounds on the performance gap between reduced-order and full-order filtering.

%Our method is based on following rationale: the sequential pipeline (``first identify a reduced order model to fit the data, then design the filter") tends to give a worse filter than a joint, filtering-oriented identification of the model wherein the `filtering-objective' is considered in the model training. 
%This is due to the fact that, in the sequential approach, the  dynamics are optimized for a different objective (in particular data-fitting) rather than the  filtering objective.%This work differs from the contribution \cite{KOELEWIJN2024} as the latter does not consider \emph{filtering-oriented} model identification for the eventual filtering task.
%As our your goal is to achieve best filtering/observer accuracy, the sequential approach in \cite{KOELEWIJN2024} will typically be less accurate, because it’s not optimized for the filtering objective. 
%The joint, filtering-oriented training proposed in this work is specifically designed to obtain filter-oriented models. 

%\MM{The paper is organized as follows:}

\section{PROBLEM SETTING}
 Consider the  discrete-time  nonlinear dynamical system:
\begin{subequations}\label{eq:system}
   \begin{align}
    x_{k+1} &= f(x_k, u_k) + w_k, \\
    y_k &= g(x_k,u_k) +v_k,
\end{align} 
\end{subequations}
where $x_k \in \R^{n_x}$ denotes the state, $y_k \in \R^{n_y}$  the measured output, and $u_k \in \R^{n_u}$ is the control input at time step $k$. The process and measurement noises, $w_k \in \R^{n_x}$ and $v_k \in \R^{n_y}$, are assumed to be independent, identically distributed (i.i.d.) white noise sequences with finite variances. 

We consider the setting where the system  \eqref{eq:system} is  \emph{high-dimensional}   (typically, having  state dimension $n_x>50$) and $n_x \gg n_y$. Our objective is to design a \emph{filter} for \eqref{eq:system} to estimate the state $x_k$ from the measurement history $(u_{0:k},y_{0:k})$. For such high-dimensional systems, implementing an extended Kalman filter or a nonlinear observer  using a \emph{full-order} model is often computationally prohibitive, particularly in resource-constrained or embedded applications.  
To address this challenge, we seek a \emph{reduced-order model} for the  system \eqref{eq:system} that is well-suited for online state estimation via filtering algorithms. 
To this end, the objective of this work is  twofold: 
\begin{itemize}
    \item[(i)] Learn a \emph{reduced-order} \emph{linear time-invariant} (RO-LTI)  state-space model with states $z_k \in \R^{n_z}$, where  $n_z\ll n_x$,  together with a state-lifting map $\Phi: \R^{n_z} \rightarrow \R^{n_x}$ that reconstructs the full-order state  from the latent state of the LTI model, \emph{i.e.}, $x_k \simeq \Phi(z_k)$.  
    \item[(ii)] Learn the RO-LTI model such that it is tailored for downstream filter design; that is, the model is optimized to enable accurate state estimation of the high-dimensional state \(x_k\), rather than merely describing the observed data.
    %the LTI model is to be tailored towards obtaining an optimal filter/observer to reconstruct the true high-dimensional state $x_k$, rather than merely describing the data generated from \eqref{eq:system}.  
\end{itemize}

We refer to this approach as \emph{filtering-oriented} reduced-order model identification.  
The problem addressed in this work is formalized in the following.

%\begin{problem}
Consider  the  RO-LTI state-space model:
\begin{subequations}\label{eq:LTImodel}
    \begin{align}
        z_{k+1} &=\! Az_{k} + Bu_k \!+\! \tilde{w}_k, \ \tilde{w}_k \!\sim\! \mathcal{N}(0,Q), \label{eq:LPVmodela}\\
        y_k &=\! Cz_k \!+\! Du_k \!+\! \tilde{v}_k, \ \tilde{v}_k \sim \mathcal{N}(0,R),
    \end{align} 
    \end{subequations}
where $z_k \in \R^{n_z}$ is the reduced-order state with $n_z  \ll n_x$. The noise terms $\tilde{w}_k \in \R^{n_z}, \tilde{v}_k \in \R^{n_y}$ are zero-mean Gaussian distributed with covariance matrices $Q, R \succeq 0 $, respectively.  Let $\theta = \{  A, B, C, D, Q, R \} \in \R^{n_{\theta}}$ denote the parameters of the RO-LTI model \eqref{eq:LTImodel}. 
In the downstream filtering task, we first  estimate the latent reduced-order state $z_k$ using a Kalman filter applied to \eqref{eq:LTImodel}. In particular, let
\begin{equation*}
    \mu_{k|k} := \mathbb{E}[z_k|y_{0:k}, u_{0:k}],
\end{equation*}
denote the posterior mean of $z_k$, where the subscript  $({\cdot|k})$ indicates conditioning on all measurements up to time  $k$.
%$\hat{x}_k \in \R^{n_x}$ of the true state by running a Kalman Filter with the reduced-order model \eqref{eq:LTImodel} to estimate the reduced dimensional state $\mu_{k|k} := \mathbb{E}[z_k|y_{0:k}, u_{0:k}]$, where the subscript in $\mu_{k|k}$ denotes the estimate computed at $k$ by considering all data up to time $k$. 

The estimate of the full-order state $\hat{x}_k \in \R^{n_x}$ is then obtained by mapping the filtered latent state $\mu_{k|k}$ through the lifting map $\Phi(\cdot)$. Specifically, 
\begin{subequations}\label{eq:reducedFilt}
    \begin{align}
    \mu_{k|k} &= \mathtt{KF}(\theta,  y_{0:k}, u_{0:k}), \label{eq:KalmanFilter} \\
    \hat{x}_k &= \Phi(\mu_{k|k}), \label{eq:lifting_map}
\end{align}
\end{subequations}
where $\mathtt{KF}(\cdot)$ denotes the recursive \emph{Kalman filter} (KF)  that computes the posterior mean  $\mu_{k|k}$ (and  covariance $P_{k|k}$) of the  RO-LTI model's state $z_k$ given the model parameters $\theta$  and the observations $(u_{0:k}, y_{0:k})$. Then,  $\Phi(\cdot)$  maps the reduced-order  state estimate $\mu_{k|k}$ to the original high-dimensional state $\hat{x}_k \in \R^{n_x}$.

Suppose we are given a dataset $\mathcal{D}= \{x_k,y_k,u_k\}_{k=0}^{N-1}$ consisting of $N$ state-output-input samples generated from the nonlinear system~\eqref{eq:system}. We now state the problem. 
\begin{problem} \label{prob}
Given a dataset $\mathcal{D}= \{x_k,y_k,u_k\}_{k=0}^{N-1}$ generated from \eqref{eq:system},  estimate the  parameters $\theta$ of the RO-LTI model \eqref{eq:LTImodel} and learn a state-lifting map $\Phi(\cdot)$  in \eqref{eq:lifting_map} such that the reconstructed state estimate  $\hat{x}_k$ in \eqref{eq:reducedFilt} matches as closely as possible to the true high-dimensional state $x_k$. \hfill $\blacksquare$
\end{problem}

We thus seek the parameters $\theta$ of a reduced-order model \eqref{eq:LTImodel}, including both the state-space matrices $(A,B,C,D)$ and the noise covariances $(Q,R)$, that are explicitly tailored to the downstream state estimation task.
Once the model parameters $\theta$ and the lifting map $\Phi(\cdot)$ are learned by solving Problem~\ref{prob}, they are used for  filtering on unseen data. %Specifically, given a new sequence of input-output measurements $u_{0:k},y_{0:k}$ (not used during training), we apply the reduced-order KF in \eqref{eq:reducedFilt} to compute an estimate $\hat{x}_k$ of the high-dimensional state using only input-output data, without requiring access to  the full high-dimensional model.
\begin{remark}
The full state $x_k$ is assumed to be available only during training, for the purpose of learning the RO-LTI model and the map $\Phi$. During inference (filtering on unseen data), only input-output measurements are available, and the state is first estimated using the reduced-order Kalman filter (eq. \eqref{eq:KalmanFilter}) and then mapped to the original space via the lifting map $\Phi$ (eq. \eqref{eq:lifting_map}). \hfill $\blacksquare$
\end{remark}

\section{METHODOLOGY}

\subsection{Learning the State-Lifting Map via Autoencoders}

We aim to learn a state-lifting map $\Phi: \R^{n_z} \rightarrow \R^{n_x}$ as introduced in \eqref{eq:lifting_map}. To this end, we employ an \emph{Autoencoder} (AE) architecture~\cite{Hinton2006} consisting of an \emph{encoder}   $\mathsf{Enc}_{\psi}: \R^{n_x} \rightarrow \R^{n_z}$ and a \emph{decoder}  network $\mathsf{Dec}_{\phi}: \R^{n_z} \rightarrow \R^{n_x}$. The weights  of the encoder/decoder networks are denoted by $\psi$ and $\phi$, respectively. The encoder maps the high-dimensional state to a latent low-dimensional state, which is subsequently lifted back to the original space via the decoder: 
\begin{equation}\label{eq:AE}
    \hat{z}_k = \mathsf{Enc}_{\psi}(x_k), \quad  \hat{x}_k = \mathsf{Dec}_{\phi}(\hat{z}_k).
\end{equation}

The decoder $\mathsf{Dec}_{\phi}(\hat{z}_k)$ serves as a parametric  model of the lifting map $\Phi(\cdot)$. Thus, learning $\Phi$ translates to computing the decoder parameters $\phi$. 
%The AE weights $(\psi, \phi)$ 
%are learned by minimizing the 
We consider the following AE
\emph{reconstruction loss},
\begin{align}\label{eq:loss_ae}
    \mathcal{L}_{\mathrm{ae}}(\psi, \phi) = \frac{1}{N}\sum \limits_{k=0}^{N-1} \|x_k - \mathsf{Dec}_{\phi}(\mathsf{Enc}_{\psi}(x_k)) \|^2_2.
\end{align}
%the AE is trained jointly with the RO-LTI model \eqref{eq:LTImodel}, such that the learned latent representation is  also suitable for downstream filtering.
This loss is used to learn a lower-dimensional latent representation of the state, which is particularly effective when the components of $x_k$ are  correlated.  
In the following, by considering the loss in \eqref{eq:AE} together with other filtering-oriented objectives, 
the AE parameters $(\psi, \phi)$ are 
jointly learned with the RO-LTI model \eqref{eq:LTImodel} parameters $\theta$.
%trained jointly  with the RO-LTI model \eqref{eq:LTImodel} to learn the parameters $\theta$  tailored for a filtering objective.

\subsection{Learning the Reduced-Order  LTI Model}

We aim to learn the parameters $\theta$ of the RO-LTI model \eqref{eq:LTImodel}, such that it is well-suited  for the downstream  filteing task.

\subsubsection{Positive Semi-definite Parameterization  Covariances}

The covariance  matrices $Q \in \R^{n_z \times n_z}, R \in \R^{n_y \times n_y}$ are parameterized via their Choleksky factorizations:
\begin{align}\label{eq:chol}
    Q = L_Q L^{\top}_Q, \   R = L_R L^{\top}_R,
\end{align}
where $L_Q$ and $L_R$ are lower-triangular matrices treated as decision variables. This parameterization 
ensures that $Q, R \succeq 0$  by construction. 
%We next formulate a loss function to learn parameters  $\theta =  \{ A, B, C, D, Q, R \}$ of the RO-LTI model \eqref{eq:LTImodel}, such that it is tailored for filtering. 
We next recall the  Kalman filter   associated with the RO-LTI model, which will be used to define the learning objective to identify the parameters $\theta =  \{ A, B, C, D, L_Q, L_R \}$.

\subsubsection{Kalman Filter Recursions}
The Kalman filter is initialized with $\mu_{0|0} = \hat{z}_0= \mathsf{Enc}_{\psi}(x_0)$, where $\hat{z}_0$ is the encoded initial state and let $P_{0|0} =P_0$ be the chosen initial latent state covariance. 

The KF recursions are given by: 
\begin{subequations}\label{eq:KF_algorithm}
    \begin{align}
 \text{prediction:}\quad   &\left\{\begin{aligned}
                \mu_{k|k-1} &= A\mu_{k-1|k-1} + B u_k, \label{eq:pred_mean}\\
P_{k|k-1} &= A P_{k-1|k-1} A^\top + Q,  
    \end{aligned}\right.\\
r_k &= y_k - (C\mu_{k|k-1} + D u_k), \label{eq:residual}\\
S_k &= C P_{k|k-1} C^\top + R, \label{eq:innov_cov}\\
\text{update:}\quad&\left\{\begin{aligned}
   \mu_{k|k} &= \mu_{k|k-1} + K_kr_k, \label{eq:post_mean}\\
P_{k|k} &= (I - K_k C)P_{k|k-1},   
\end{aligned}\right.
\end{align}
\end{subequations}
where the Kalman gain is given by 
$K_k = P_{k|k-1} C^\top S_k^{-1}$.
Here, $\mu_{k|k-1} \in \R^{n_z}, P_{k|k-1} \in \R^{n_z \times n_z}$ in \eqref{eq:pred_mean} denote the \emph{predicted} mean and covariance of the latent state $z_k$,  respectively. The term $r_k \in \R^{n_y}$ in \eqref{eq:residual} is the innovation residual and  $S_k \in \R^{n_y \times n_y}$ in \eqref{eq:innov_cov} is the innovation covariance. The filter's posterior update based on the output measurements $y_k$ is given in \eqref{eq:post_mean} where $\mu_{k|k}$ and $P_{k|k}$ denote the updated mean and covariance of the state $z_k$, respectively.  Note that the posterior mean $\mu_{k|k} = \mathbb{E}[z_k|y_{0:k}, u_{0:k}]$ given by the
algorithm \eqref{eq:KF_algorithm}, depends on the model parameters $\theta$ and the measurements $(y_{0:k}, u_{0:k})$ up to time $k$. We denote this dependence compactly as, 
\begin{equation}\label{eq:mean_estimate}
    \mu_{k|k}(\theta) = \mathtt{KF}(\theta, y_{0:k}, u_{0:k}),
\end{equation}
where  $\mathtt{KF}(\cdot)$ represents the recursion in \eqref{eq:KF_algorithm}.

Based on the above formulation, we learn the model parameters $\theta$ by optimizing objectives tailored to the filtering task. To this end,  we combine  (i) the Kalman filter negative log-likelihood  and (ii) a  state reconstruction loss  that enforces the state estimate given by the KF, mapped into the original x-space via the decoder $\hat{x}_k = \mathsf{Dec}_{\phi}(\mu_{k|k}(\theta))$ to match the true states. In particular,
we consider the following loss functions:
\begin{align}%\label{eq:loss_filter}
\mathcal{L}_{\mathrm{nll}}(\theta) &= \frac{1}{2N} \sum \limits_{k=0}^{N-1} \Big(
\log\det S_k + r_k^\top S_k^{-1}r_k
\Big)  \label{eq:loss_nll}\\
    \mathcal{L}_{\mathrm{filter}}(\theta, \phi) &=   \frac{1}{N}\sum \limits_{k=0}^{N-1} \|x_k - \mathsf{Dec}_{\phi}(\mu_{k|k}(\theta))\|^2_2 \label{eq:loss_filter}
\end{align}

The loss $\mathcal{L}_{\mathrm{nll}}(\theta)$ in \eqref{eq:loss_nll} is the \emph{negative log-likelihood} loss of the Kalman filter, computed from the innovation residuals $r_k$ in \eqref{eq:residual} and their covariance $S_k$ in \eqref{eq:innov_cov}. This term encourages  the learned model  to be consistent with the measurements $y_{0:N}$. The loss $\mathcal{L}_{\mathrm{filter}}(\theta,  \phi)$ in \eqref{eq:loss_filter} minimizes the mismatch between the true state $x_k$ and the reconstructed estimate obtained by decoding the filtered latent mean as $\mathsf{Dec}_{\phi}(\mu_{k})$. This term explicitly promotes accurate state estimation in the original high-dimensional space.

\subsection{Latent Space Alignment}

The latent representation $\hat{z}_k$ obtained from the autoencoder in \eqref{eq:AE} is, in general,  not  aligned with the latent state of the RO-LTI model \eqref{eq:LTImodel}.  To enforce consistency between these representations, we introduce the following latent alignment loss:
\begin{align}\label{eq:loss_latent}
\mathcal{L}_{\mathrm{latent}}(\phi, \theta) = \frac{1}{N}\sum \limits_{k=0}^{N-1} \| \hat{z}_k - \mu_{k|k}(\theta) \|_2^2.
 \end{align}
%This loss aims at minimizing the mismatch between the encoded latent state $\hat{z}_k$ and  the filtered mean state estimate $\mu_{k|k}(\theta)$ obtained from the Kalman filter \eqref{eq:mean_estimate}. 
Intuitively, if the autoencoder learns a meaningful low-dimensional representation of the underlying system state, then $\hat{z}_k$ serves as a reference latent state trajectory. %The alignment loss encourages the RO-LTI model and the Kalman filter to operate in a latent space consistent with this representation.

%Note that if the autoencoder learns the `true' underlying lower order state, then $\hat{z}_k$ serve as the true latent reference states for the Kalman filter and the objective \eqref{eq:loss_latent} helps the model learning towards optimal parameters for filtering. 

\subsection{Filtering-oriented Learning of the RO-LTI Model}

 We jointly learn the RO-LTI model parameters $\theta$ and the AE parameters $\psi, \phi$  
 by minimizing the composite objective:
\begin{align}\label{eq:loss_total}
    \mathcal{L}(\theta,\psi,\phi) = &\lambda_{\mathrm{ae}} \mathcal{L}_{\mathrm{ae}} + \lambda_{\mathrm{nll}} \mathcal{L}_{\mathrm{nll}}+ \lambda_{\mathrm{filt}} \mathcal{L}_{\mathrm{filt}} +  \lambda_{\mathrm{latent}} \mathcal{L}_{\mathrm{latent}}, 
\end{align}
where $\lambda_{\mathrm{ae}}, \lambda_{\mathrm{nll}}, \lambda_{\mathrm{filt}}, \lambda_{\mathrm{latent}}   \geq 0$ are the tuning hyperparameters that balance the  contribution of each term.

The composite loss \eqref{eq:loss_total} embodies the \emph{task-oriented} learning idea of the proposed approach. In summary, the term $\mathcal{L}_{\mathrm{nll}}$  shapes the LTI model parameters and noise covariances through the Kalman filter likelihood, ensuring that the reduced-order model is optimized not merely for system approximation, but explicitly for \emph{filtering performance}. The term $\mathcal{L}_{\mathrm{filt}}$  reinforces this by penalizing reconstruction error in the original state space after decoding, directly linking the latent filtering objective to the estimation quality in the original state space. The terms $\mathcal{L}_{\mathrm{ae}}$
 and $\mathcal{L}_{\mathrm{latent}}$ ensure geometric consistency of the latent space and alignment between the encoder outputs and the Kalman filter states, respectively. Critically, all terms are optimized simultaneously: the reduced-order model, noise covariances, and autoencoder parameters are learned jointly rather than sequentially. This is in contrast to the classical two-stage paradigm, in which a reduced-order model is first identified for system approximation, and a filter is subsequently designed as a separate step. In the proposed formulation, no such separation exists-the reduced-order model is the filtering model, and its parameters are shaped entirely by the estimation task.

Once the  model and AE are trained, at the inference stage, we reconstruct the estimate of true state as described in \eqref{eq:reducedFilt}. Note that only the reduced-order Kalman filter and the decoder are required, while the encoder is not needed

\begin{remark}[Differentiation through the Kalman filter]
    The Kalman filter recursions in \eqref{eq:KF_algorithm} define a differentiable computational graph with respect to the model parameters $\theta$. In particular,  the loss functions \eqref{eq:loss_nll}, \eqref{eq:loss_filter}, and \eqref{eq:loss_latent} depend on $\mu_k(\theta), r_k, S_k$ which are obtained via the recursive  filtering equations. Automatic differentiation therefore applies the chain rule through the entire sequence of recursions, enabling the computation of the gradient of the total loss $\mathcal{L}(\theta,\psi,\phi)$ with respect to $\theta$. Standard gradient-based optimizers (e.g., Adam or SGD) are then used to update the parameters $\theta, \psi, \phi$. \hfill $\blacksquare$
\end{remark}

\begin{remark}[Reduced-order System Identification]\label{rem:SysID}
In a conventional system identification (SID) setup, we  learn a reduced-order model by minimizing  a \emph{prediction-based} objective of the form:
\begin{align}\label{eq:SysID}
    \mathcal{L}_{\rm SID}(\theta, \psi, \phi) =& \lambda_{\rm ae} \mathcal{L}_{\rm ae} + \frac{1}{N} \sum \limits_{k=0}^{N-1}\|x_{k+1}- \mathsf{Dec}_{\phi}(\tilde{z}_{k+1}) \|^{2}_2 \nonumber  \\
    &+ \frac{1}{N} \sum \limits_{k=0}^{N-1}\|\tilde{z}_{k+1}- \hat{z}_{k+1} \|^{2}_2 \nonumber \\
  \mathrm{s.t.}  \quad \hat{z}_k &= \mathsf{Enc}_{\psi}(x_k), k =0,\ldots,N-1, \nonumber \\ 
  \tilde{z}_{k+1} &= A \hat{z}_k + B u_k, y_k = Cz_k + Du_k.
  \end{align}
 This corresponds to jointly learning a reduced-order LTI model and a lifting map based on reconstruction and one-step prediction errors, similar to the approaches introduced in        \cite{KOELEWIJN2024, MASTI2021}. A key distinction from the proposed approach is that \eqref{eq:SysID} does not incorporate any filtering objective. As shown  in the  case study (Section~\ref{sec:case_study}), models learned via \eqref{eq:SysID} yield inferior filtering performance compared to the proposed filtering-oriented formulation. \hfill $\blacksquare$ %This is potentially also due to the fact that the covariance matrices $Q,R$ required for the KF are not available and have to be chosen by trial-and-error. On the other hand, in our approach the covariances are part of the model parameters which are learned during the training.  
\end{remark}

\section{Generalization Bounds on Filtering Performance} \label{sec:sub-opt}

The goal of  this section is to establish probabilistic  bounds that quantify the performance gap between   filtering  with the proposed \emph{reduced-order}   and a \emph{full-order} model.  %In particular, for a new unseen test dataset,  we aim to characterize how much filtering performance is degraded when using the reduced-order model by establishing probabilistic upper bounds on this gap. 
Our analysis builds on the results developed in \cite{BUSETTO2025}  by some of the authors, which are adapted here to the reduced-order filtering setting.

\subsection{Definitions}
Let $\thm \in \Theta \subset  \R^{n_{\thm}}$ denote both the parameters of the reduced-order latent LTI  model $\theta$ and the decoder  parameters $\phi$. Let $\ths \in \Theta^{\star} \subset  \R^{n_{\ths}}$ denote the parameters of the full-order state-space model of the system \eqref{eq:system}, including the process and measurement noise covariances. 

Let us denote the state estimates obtained from the reduced-order and full-order Kalman filters as $\hat{x}_k(\thm, y_{0:k}) = \mathsf{Dec}_{\phi}(\mathbb{E}[z_k| \theta, y_{0:k}])$ and $\hat{x}^{\star}_k(\ths, y_{0:k}) = \mathbb{E}[x_k| \ths, y_{0:k}]$, respectively\footnote{The dependence on input sequence $u_{0:k}$ is dropped for brevity.}.    
To quantify estimation performance over a trajectory of
length $T$, we define the root-mean-square error (RMSE) as,
%Corresponding to these state estimates, let
%$f: \R^{Tn_{x}} \rightarrow \R_{\geq0}$  be the  state-estimation error  defined as the following RMSE:
\begin{equation}\label{eq:per_rmse}
   f(x,\hat{x}(\bar{\theta}, y_{0:T-1})) = \sqrt{\frac{1}{Tn_x}\sum \limits_{k=0}^{T-1} \| x_k - \hat{x}_k(\bar{\theta}, y_{0:T-1}) \|_{2}^{2}}, 
\end{equation}
where  $\bar{\theta}$ denotes  either the RO-LTI/decoder parameters $\thm$ or the  true full-order parameters $\ths$. Here, we denote  the true state vector sequence as $x :=\left[x^{\top}_0 \cdots x^{\top}_{T-1} \right] \in R^{Tn_x} $ and 
$\hat{x}(\bar{\theta},y_{0:T-1}) :=\left[\hat{x}^{\top}_{0}(\bar{\theta},y_{0}) \cdots \hat{x}^{\top}_{T-1}(\bar{\theta},y_{0:T-1}) \right] \in R^{Tn_x} $ denotes the corresponding sequence of estimated states.
 
 We define the performance gap between reduced-order and
full-order filtering as  follows:
\begin{align} \label{eq:perc_gap}
\Psi(\thm, \ths|x,y) \!:=\! 
f(x,\hat{x}(\thm,y_{0:T-1})) \!-\! f(x,\hat{x}^{\star}(\ths), y_{0:T-1}).
\end{align}

 Next, we formalize the data-generating process.
For the data-generating system in \eqref{eq:system}, let $p(u), p(x_0), p(w)$ and $p(v)$ denote the probability density functions over the inputs, initial state, process noise and measurement noise, respectively. Let $p(\mathcal{D}) := p(u) \otimes p(x_0) \otimes p(w) \otimes p(v)$ be the product measure  denoting the data-generating distribution. Let $\{(u_{0:T}^{(i)}, x^{(i)}_0, w_{0:T}^{(i)}, v_{0:T}^{(i)} ) \}_{i}^{m}$ denote a validation set of $m$ independent realizations from $p(\mathcal{D})$ and let    $\{(x_{0:T}^{(i)}, y_{0:T}^{(i)} )\}_{i=1}^{m}$ be the corresponding states and measured outputs. In other words, the set $\{(x_{0:T}^{(i)}, y_{0:T}^{(i)}) \}_{i=1}^{m}$ denotes  $m$ independent validation datasets, where the $i$-th validation set contains $T$ samples of states and outputs generated by exciting the system \eqref{eq:system} with the input sequence $u^{(i)}_{0:T-1}$, having an initial condition $x^{(i)}_0$ and the corresponding noise realization affecting the system are $w^{(i)}_{0:T}, v^{(i)}_{0:T}$.  We write  $\{(x_{0:T}^{(i)}, y_{0:T}^{(i)}) \}_{i=1}^{m}$ compactly  as $\{(x^{(i)}, y^{(i)}) \}_{i=1}^{m}$.

For each trajectory, we define the corresponding estimation
error and performance gap,
%The validation set $\{(x^{(i)}, y^{(i)}) \}_{i=1}^{m}$ can thus corresponds to $m$ realizations of the state RMSE and performance gap $\Psi$, 
namely, for $i=1,\dots,m$, 
\begin{align}
f^{(i)} := f(x^{(i)}, x(\bar{\theta},y^{(i)})), 
\quad  
\Psi^{(i)} := \Psi(\thm,\ths|x^{(i)},y^{(i)}). 
%\quad i=1,\dots,m.
\end{align}

Let $(x^{\rm new},y^{\rm new})$ denote an independent trajectory generated from the underlying distribution
$p(\mathcal{D})$, not used during training or validation, 
with corresponding state-estimation error $f^{\rm new}$ and the performance gap $\Psi^{\rm new}$.

Finally, let $\mathbb{P}_\Psi$ be the   probability measure  of the random variable $\Psi$. 
 Similarly, let $S=\{\Psi^{(i)}\}_{i=1}^m$ denote the validation samples and $\mathbb{P}_S$ 
denote the probability distribution of $S$. 

We now state a result that provides a probabilistic bound on the performance gap.

\begin{theorem}[Generalization bounds~\cite{BUSETTO2025}] \label{theorem:gap}
For any $\alpha,\delta \in (0,1)$, define
\[
k^\star := \Big\lceil m\,(1-\alpha+\varepsilon_m)\Big\rceil, 
\qquad 
\varepsilon_m := \sqrt{\frac{\log(2/\delta)}{2m}}.
\]
Assume that the number of validation samples $m$ is  sufficiently large such that  $\alpha \geq  \varepsilon_m$ (equivalently,  $k^\star \le m$).  
Then, with probability at least $1-\delta$  over the  validation set 
$S=\left\{\Psi^{(i)} \right\}_{i=1}^{m}$, it holds that
\begin{align} \label{eqn:Psi_probl}
\mathbb{P}_{\Psi}\!\left(\Psi \le \Psi_{(k^\star)}\right) \;\ge\; 1 - \alpha,
\end{align}
where  $\Psi_{(k)}$ denotes the $k$-th 
order statistic of the validation values, i.e., 
\begin{align}
\Psi_{(1)} \le \Psi_{(2)} \le \dots \le \Psi_{(m)},
\end{align}
with $\Psi_{(k)}$ being the $k$-th smallest value in the ordered sequence.  
Equivalently,
\begin{align} \label{eqn:Psi|S_probl}
\mathbb{P}_S\Big( \, \mathbb{P}_{\Psi}\big(\Psi 
   \le \Psi_{(k^\star)} \,\big|\, S\big) \;\ge\; 1-\alpha \,\Big) \;\ge\; 1-\delta.
\end{align}
\hfill $\blacksquare$
\end{theorem}
We refer to \cite[Theorem 1]{BUSETTO2025} for details and proof.

Intuitively,  
 for any  new independent trajectory $(x^{\mathrm{new}}, y^{\mathrm{new}}) \sim p(\mathcal{D})$ with corresponding gap $\Psi^{\mathrm{new}}$, Theorem~\ref{theorem:gap}  guarantees that
\begin{align}
\mathbb{P}_\Psi\!\left(\Psi^{\mathrm{new}} \le \Psi_{(k^\star)} \,\middle|\, S\right) \;\ge\; 1-\alpha,
\end{align}
and this statement holds with probability at least $1-\delta$ with respect to the randomness of the validation set $S$.
In other words, $\Psi_{(k^\star)}$ serves as a distribution-free upper bound on the performance degradation incurred by the reduced-order filter on unseen data. This bound  does not require any assumptions on the underlying data distribution.

\section{Case Study}
\label{sec:case_study}

We demonstrate the proposed framework on a high-dimensional system arising from partial differential equations (PDEs), which are common in applications such as fluid dynamics, structural mechanics, and materials science. In particular, we consider a nonlinear heat diffusion problem in a one-dimensional rod. The state represents the temperature at 100 spatial locations along the rod, resulting in a system with state dimension $n_x\!=\!100$.

\subsection{Heat diffusion in a metal rod}
\label{ssec:benchmark}

A central objective in heat conduction analysis  is to determine the temperature distribution along the rod under given boundary conditions \cite{bergman2016fundamentals}. We consider a 1-D heat conduction problem in a steel rod of length  $L=10$ cm. Let $\T(\mathsf{x}, t)$ denote the temperature along the rod at position $\mathsf{x} \in [0 \ \ L] \ \mathrm{m}$ and  time $t$, whose evolution is governed by the following   heat conduction PDE:
\begin{align}\label{eq:heat_conduction}
    \rho c \frac{\partial \T(\mathsf{x},t)}{\partial t} = \frac{\partial}{\partial \mathsf{x}} \left(\bar{k}(\T)\frac{\partial \T(\mathsf{x},t)}{\partial \mathsf{x}} \right), \   \mathsf{x} \in [0 \ L], \ t > 0, 
\end{align}
where  $\rho, c, k(\cdot)$ characterize the material  properties of the rod  whose values are set as follows: mass density $\rho\!=\!7800\,\mathrm{kg/m^3}$; specific heat capacity $c\!=\!486\,\mathrm{J/(kg\,K)}$;  temperature-dependent
conductivity $\bar{k}(\T)\!=\!k_0(1+\alpha \T) \,\mathrm{W m^{-1} K^{-1}}$ with  temperature coefficient $\alpha\!=\!-2\!\times\!10^{-4}\,\mathrm{K}^{-1}$.

A finite-element spatial discretization  at $100$   locations  $\mathsf{x}^{j} \in [0, L], j=1,\ldots, 100$,  
and a semi-implicit Euler integrator with sampling time $\Delta t\!=\!2\,\mathrm{s}$ of the PDE  \eqref{eq:heat_conduction},
 results in a stable discrete-time nonlinear state-space model of the form \eqref{eq:system} with dimensions $n_x=100, n_u =2, n_y = 5$. The state $x_k \in \R^{100}$ represents the temperatures  at the discretized spatial nodes, \emph{i.e.}, $x_k= \left[\T(\mathsf{x}^1,k) \ \cdots  \ \T(\mathsf{x}^{100},k)  \right]^{\top}$.  The outputs $y_k \in \R^{5}$ correspond to temperature measurements only at $5$ sensor locations, specifically at nodes
  $y_k = \{\T(\mathsf{x}^{i},k) \}, i=20,40,60, 80,100$.
 The process and measurement noise  are Gaussian distributed $w_k\!\sim\!\mathcal{N}(0,0.1\,I_{100})$,
$v_k\!\sim\!\mathcal{N}(0,0.1\,I_5)$. The two inputs $u_k \in \R^2$ are the time-varying Dirichlet boundary temperatures
$u_k\!=\![\T(0,k) \ \ \T(L,k)]^{\top}$ at the left-most and the right-most end of the rod. 

\subsection{Data}
The training data are generated by exciting the nonlinear heating rod system  with multi-sine boundary inputs. In particular, the left boundary  temperature $\T(0,k)$ is a multi-sine signal consisting of four
harmonics (periods between 360--7200\,s) with an amplitude of $300 \pm 50 \,^\circ\mathrm{C}$ representing furnace-like conditions.  The right boundary input temperature $\T(L,k)$ is also a multi-sine signal with four harmonics  (periods between 360--7200\,s) and amplitude of $26 \pm 4\,^\circ\mathrm{C}$, modeling heat exchange with ambient air.  The temperature difference at the two ends creates a permanent temperature gradient across the rod from $\sim 300\,^\circ\mathrm{C}$ to $\sim 25\,^\circ\mathrm{C}$.

For the test  dataset, we use    single-sinusoidal boundary inputs with
 amplitudes  $(300 \pm 25\,^\circ\mathrm{C}), (25 \pm 2\,^\circ\mathrm{C})$ and periods of $(1200,3600) s$ for the left and right boundary inputs, respectively.
The number of training and test samples is $N_{\mathrm{tr}}\!=\!N_{\mathrm{te}}\!=\!1000$. All signals are min-max normalized using training data statistics. 

Furthermore, to evaluate the probabilistic bounds on performance gap between the full-order and reduced-order filter (see Section~\ref{sec:sub-opt}), we generate  $m=1000$ independent validation datasets, each consisting of $T=500$ samples of state-input-output trajectory $\mathcal{D}_{\rm val} = \{\{x^{(i)}_k, y^{(i)}_k, u^{(i)}_k \}_{k=0}^{500} \}_{i=1}^{1000}$. For each trajectory, the boundary inputs are single-sinusoidal signals with randomly sampled amplitudes, periods, and phases, ensuring i.i.d. validation conditions.

\subsection{Model configuration and training}
For the AE in \eqref{eq:AE} and RO-LTI model \eqref{eq:LTImodel}, we set the latent state dimension to $n_z =8$. Thus, we are considering the compression of the state dimension from $x_k\!\in\!\mathbb{R}^{100}$ to
$z_k\!\in\!\mathbb{R}^{8}$. 
The encoder is a feedforward MLP with three hidden layers,  
and $\mathrm{tanh}$ activations, mapping the input to a latent representation. The decoder mirrors this architecture with the same layer sizes in reverse.

We train the AE and RO-LTI model with a latent differentiable Kalman filter by minimizing the multi-objective loss \eqref{eq:loss_total}, using the \emph{Adam} optimizer with learning rate $\eta = 10^{-2}$ over $350$ epochs. The training proceeds in three phases: (i) during the first 150 epochs, $\lambda_{\rm filt} = \lambda_{\rm latent} = \lambda_{\rm ae} = 1$ and $\lambda_{\rm nll} = 0$; (ii) over the next 100 epochs, $\lambda_{\rm filt} = \lambda_{\rm latent} = 1$, $\lambda_{\rm ae} = 0$, and $\lambda_{\rm nll}$ is linearly increased from 0 to 1; (iii) during the final 100 epochs, $\lambda_{\rm filt} = \lambda_{\rm latent} = \lambda_{\rm nll} = 1$ and $\lambda_{\rm ae} = 0$. This rationale behind the training  procedure is to emphasize learning the encoder/decoder networks  first, in the initial part of the training to guide the AE  to obtain a correct latent representation. Once this is done, the weights $\lambda_{\rm ae}$ associated with learning the encoder are set to zero and the training is focused on learning the model and the KF. Similar training heuristics have also been reported in \cite{MASTI2021}.

\subsection{Results}
\label{ssec:results}

\begin{figure}[t]
  \centering
  \includegraphics[width=\columnwidth]{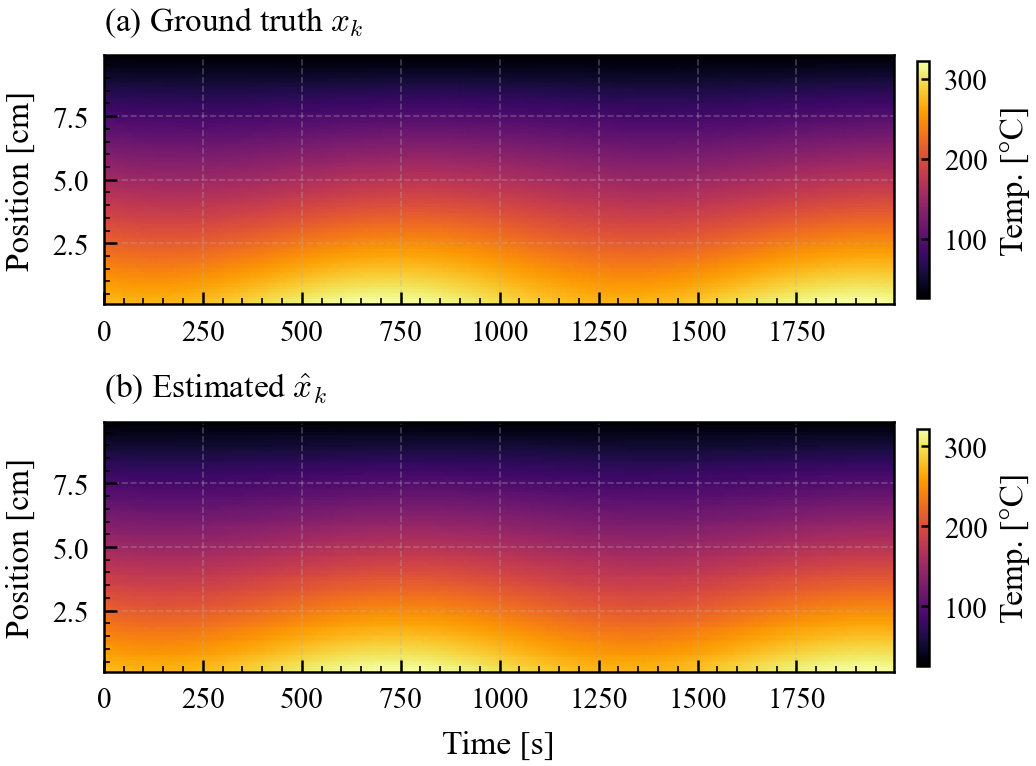}  \caption{%
  Temperature fields along the rod at different sampling times.
    (a)~Ground-truth states. (b)~ROKF-estimated states.
    }
  \label{fig:staterec}
\end{figure}

\begin{figure}[t]
  \centering
  \includegraphics[width=\columnwidth]{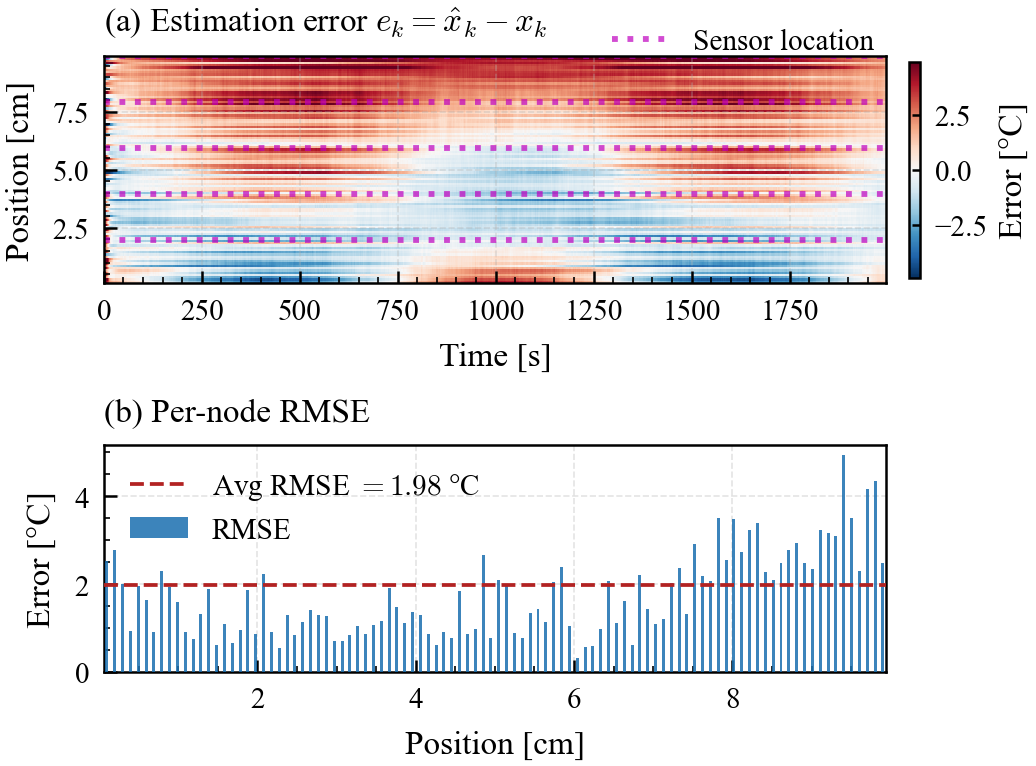}
  \caption{%
    (a)~Estimation error $x_k-\hat{x}_k$; dotted lines show the sensor location on the rod.
    (b)~Per-node RMSE (blue bars); dashed line is the
    average RMSE over $100$ nodes.
    %Errors are spatially uniform with no systematic bias.%
  }
  \label{fig:staterec_err}
\end{figure}

Figure~\ref{fig:staterec} shows the ground-truth and estimated
temperature fields $\T(\mathsf{x}^j,k)$ of the rod on the test dataset at each node location   $\mathsf{x}^j$ (y-axis) and at time $k$ (x-axis).
The values of the temperature $\T(\mathsf{x}^j,k)$ are represented by the color gradient. The figure also shows 
the  RMSE  errors for each node position $\mathsf{x}^j$. The proposed \emph{reduced-order Kalman filter}  (ROKF) is able to reconstruct the full space--time field from five
sparse observations with an average RMSE of about $2^{\circ} \mathrm{C}$ across all nodes. 
Corresponding to the temperature fields in Fig.~\ref{fig:staterec},   we also depict the state estimation errors $x_k - \hat{x}_k$ in Fig.~\ref{fig:staterec_err} over time at each node. 
We observe spatially uniform errors and no systematic bias at interior nodes.

\subsection*{Performance comparision}

%The proposed filtering approach based on RO-LTI trained with filtering-oriented objective will be termed as \emph{reduced-order Kalman filter} (ROKF).
We compare the proposed ROKF 
against: (i) a full-order extended Kalman filter (FOEKF) using a true full-dimensional system model \eqref{eq:system} and (ii) a system identification based KF (SIDKF), where first a RO-LTI model is obtained from data as detailed in Remark~\ref{rem:SysID}, without considering a differentiable Kalman filter in the architecture,  minimizing the loss $\mathcal{L}_{\rm SID}$ as given in \eqref{eq:SysID}. 

In the FOEKF, we consider an EKF with full-state model $\bm{x}_k\!\in\!\mathbb{R}^{100}$, computing analytic Jacobians of the true model \eqref{eq:system}
and the true noise covariances, serving as the \emph{oracle}. The FOEKF is also used to compute the performance gap $\Psi(\cdot)$ in \eqref{eq:perc_gap} w.r.t. to the proposed RO-LTI model filter. 

For SIDKF training, an identical encoder--decoder architecture and latent dimension ($n_z\!=\!8$) as in the proposed model is considered. At the inference, we run  a Kalman filter  with the learned reduced-order model by setting the covariance matrices to $Q = 0.01I_{8}, R=0.01I_{5}$,  chosen by trial-and-error. We recall that the model in SIDKF is trained \emph{without} the Kalman filter in the loop and thus, $Q, R$ are not optimized during the training. 

\begin{figure}[t]
  \centering
  \includegraphics[width=\columnwidth]{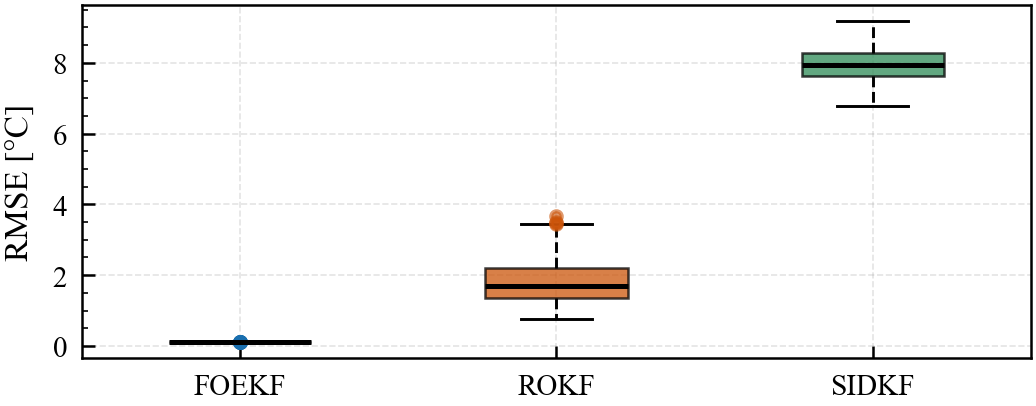}
  \caption{%
  Performance comparison on test sets:
     RMSE computed for  $1000$ Monte-Carlo trials for full-order (FOEKF), proposed reduced-order (ROKF), and system identification-based (SIDKF) Kalman filters.
 }
  \label{fig:boxplots}
\end{figure}

We compare the ROKF filtering performance quantified in terms of the RMSE as in \eqref{eq:per_rmse} with FOEKF and SIDKF. The RMSEs are computed
for $m\!=\!1000$ independent Monte-Carlo runs, each with a new
noise realization and independently randomized inputs.
Fig.~\ref{fig:boxplots} shows the boxplots of the RMSEs for the three approaches.
The FOEKF achieves the lowest RMSE as expected.
The ROKF substantially outperforms the SIDKF method, indicating that
jointly optimizing the model and the decoder with the Kalman filter
objective yields a latent model representation better matched to the state estimation task than one obtained by minimizing fitting error
alone. 

\subsection*{Probabilistic  bounds on the performance gap}
We now empirically test the probabilistic bounds derived in Theorem~\ref{theorem:gap} on the performance gap  between the FOEKF and the proposed ROKF.  We set $\alpha\!=\!0.05$, $\delta\!=\!0.05$ and  consider $m=1000$ independent validation samples. For each validation sample $i=1,\ldots,1000$, we compute the performance gap $\Psi^{(i)}$ between FOEKF and ROKF, as defined in  \eqref{eq:perc_gap} and order the set $\{\Psi^{(i)} \}_{i=1}^{1000}$ as $\Psi_{(1)} \leq \cdots \leq \Psi_{(1000)}$.   
By applying Theorem~\ref{theorem:gap}, we obtain
$\varepsilon_{1000}\!=\!0.0429$ and the bound index
$k^\star\!=\!993$. This corresponds to certified theoretical bound on the performance gap to be $\Psi_{(k^\star)}\!=\!3.23\,^\circ\mathrm{C}$. In other words, 
the theoretical bound implies that $\mathbb{P}_\Psi(\Psi \leq 3.23) \geq 0.95$ for a gap $\Psi$ computed over a fresh test set, with probability  at least $1-\delta$ over the validation set. 

\begin{figure}[t]
  \centering
\includegraphics[width=\columnwidth]{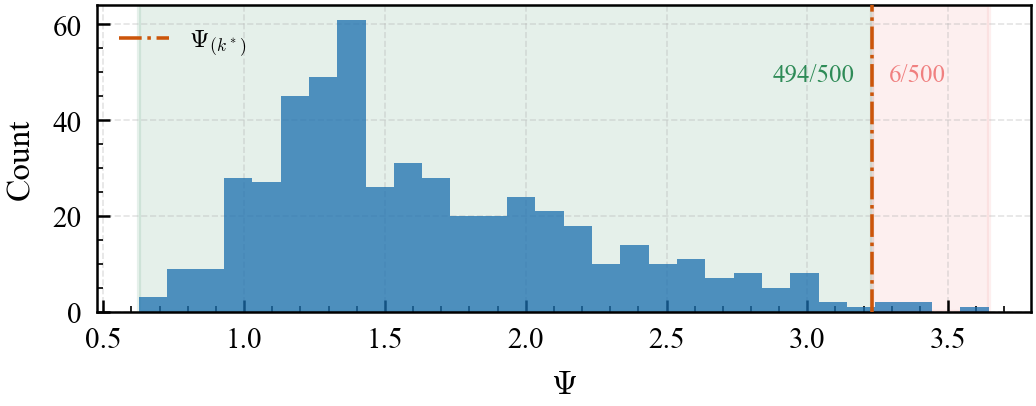}
  \caption{%
    Histogram of the performance gap $\Psi$ computed from $500$ independent test sets. The dash-dotted red line indicates the theoretical bound $\Psi_{(k^\star)} = 3.23\,^\circ\mathrm{C}$ (Theorem~\ref{theorem:gap}). The shaded green region corresponds to performance gaps below the bound $\Psi \leq \Psi_{(k^\star)}$ ($494$ out of $500$), while the shaded pink region represents those exceeding the bound ($6$ out of $500$).
  }
  \label{fig:psi_hist}
\end{figure}

To empirically validate the bound, we generate 500 independent test sets and compute the corresponding performance gaps $\{\Psi^{(j)}\}_{j=1}^{500}$. We observe that $98.8\%$ (494 out of 500) of these satisfy $\Psi^{(j)} \le \Psi_{(k^\star)}$, thus satisfying the theoretical guarantee of $1 - \alpha = 0.95$. %This empirical result is therefore consistent with Theorem~\ref{theorem:gap}. 
Figure~\ref{fig:psi_hist} shows the histogram of the test performance gaps together with the certified bound $\Psi_{(k^\star)} = 3.23\,^\circ\mathrm{C}$. Summarizing,  with $1-\delta = 0.05$ confidence over the draw of the validation set, the performance gap between the FOEKF and ROKF will be below
$3.23\,^\circ\mathrm{C}$ for a   
a  new test sample with at least $1-\alpha = 0.95$ probability.

%\begin{figure}[t]
%  \centering
%\includegraphics[width=\columnwidth]{fig3_theorem_cdf.png}
%  \caption{%
%    Empirical CDF of $\Psi\!=\!\mathrm{RMSE}_{\mathrm{FOEKF}}-%\mathrm{RMSE}_{\mathrm{ROKF}}$ computed from $m\!=\!1000$ validation sets. The bounds on %the performance gap for  $\alpha\!=\!0.05$, $\delta\!=\!0.05$.
%    Dashed green: empirical $(1\!-\!\alpha)$-quantile
%$\Psi_{(k_{\mathrm{emp}})}\!=\!2.80\,^\circ\mathrm{C}$;
%    dash-dotted red: theoretical bound
%$\Psi^{(k^\star)}\!=\!3.23\,^\circ\mathrm{C}$
%    (Theorem~\ref{theorem:gap});
%    dotted horizontal: $1\!-\!\alpha$ coverage level.%
%  }
%  \label{fig:cdf}
%\end{figure}

\section{Conclusions}

We presented a learning-based framework for identifying reduced-order models tailored to state estimation in high-dimensional nonlinear systems. By jointly learning a latent linear dynamical model and a lifting map through an autoencoder, and embedding a differentiable Kalman filter within the training loop, the proposed approach enables accurate and computationally efficient state estimation entirely in a low-dimensional space. Furthermore, the integration of conformal prediction provides probabilistic guarantees on the performance gap between reduced-order and full-order filtering, without requiring assumptions on the data distribution.

Overall, this work highlights a shift from model-centric representations, such as Koopman embeddings, to task-oriented latent representations explicitly designed for filtering. Future work will investigate extensions to linear parameter-varying latent models, as well as meta-filtering approaches for learning universal estimators across classes of dynamical systems.

\addtolength{\textheight}{-12cm}   % This command serves to balance the column lengths

%%%%%%%%%%%%%%%%%%%%%%%%%%%%%%%%%%%%%%%%%%%%%%%%%%%%%%%%%%%%%%%%%%%%%%%%%%%%%%%%

%%%%%%%%%%%%%%%%%%%%%%%%%%%%%%%%%%%%%%%%%%%%%%%%%%%%%%%%%%%%%%%%%%%%%%%%%%%%%%%%

%%%%%%%%%%%%%%%%%%%%%%%%%%%%%%%%%%%%%%%%%%%%%%%%%%%%%%%%%%%%%%%%%%%%%%%%%%%%%%%%

%%%%%%%%%%%%%%%%%%%%%%%%%%%%%%%%%%%%%%%%%%%%%%%%%%%%%%%%%%%%%%%%%%%%%%%%%%%%%%%%

\bibliographystyle{plain}
\bibliography{references}

\end{document}